%% file: PRL-Agamma_v0.7.tex
\RequirePackage{lineno}
\documentclass[aps,prl,showpacs,superscriptaddress,amsmath,amssymb,amsfonts,twocolumn,floatfix]{revtex4-1}
 
\usepackage{graphicx}
\usepackage{hyperref}

\newcommand{\CP}{{\ensuremath{C\!P}}}
\newcommand{\Acp}{\ensuremath{\mathcal{A}_\CP}}
\newcommand{\stat}{\ensuremath{\mathrm{~(stat)}}}
\newcommand{\syst}{\ensuremath{\mathrm{~(syst)}}}
\newcommand{\Dbar}{\ensuremath{\overline{D}{}}}

\newcommand{\tev}{\ensuremath{\mathrm{Te\kern -0.1em V}}}
\newcommand{\gev}{\ensuremath{\mathrm{Ge\kern -0.1em V}}}	
\newcommand{\mev}{\ensuremath{\mathrm{Me\kern -0.1em V}}}	
\newcommand{\kev}{\ensuremath{\mathrm{ke\kern -0.1em V}}}	
\newcommand{\massgev}{\mbox{\gev/$c^2$}}			
\newcommand{\massmev}{\mbox{\mev/$c^2$}}			
\newcommand{\pgev}{\mbox{\gev/$c$}}				
\newcommand{\pmev}{\mbox{\mev/$c$}}				

\begin{document}

\title{Measurement of indirect \CP-violating asymmetries in $D^0\to K^+K^-$ and $D^0\to \pi^+\pi^-$ decays at CDF}
\input{author}
\date{\today}

\begin{abstract}
We report a measurement of the indirect \CP-violating asymmetries ($A_{\Gamma}$) between effective lifetimes of anticharm and charm mesons reconstructed in \mbox{$D^0\to K^+ K^-$} and \mbox{$D^0\to \pi^+\pi^-$} decays. We use the full data set of proton-antiproton collisions collected by the Collider Detector at Fermilab experiment and corresponding to $9.7$~fb$^{-1}$ of integrated luminosity. The strong-interaction decay $D^{*+}\to D^0\pi^+$ is used to identify the meson at production as $D^0$ or $\Dbar^0$. 
We statistically subtract $D^0$ and $\overline{D}^0$ mesons originating from $b$-hadron decays and measure the yield asymmetry between anticharm and charm decays as a function of decay time.  We measure $A_\Gamma (K^+K^-) = \bigl(-0.19 \pm 0.15 \stat \pm 0.04 \syst\bigr)\%$ and $A_\Gamma (\pi^+\pi^-)= \bigl(-0.01 \pm 0.18 \stat \pm 0.03 \syst\bigr)\%$. The results are consistent with the hypothesis of \CP\ symmetry and their combination yields $A_\Gamma = \bigl(-0.12 \pm 0.12)\%$.
\end{abstract}


\maketitle

%
The noninvariance of the laws of physics under the simultaneous transformations of parity and charge conjugation (\CP\ violation) is described in the standard model (SM) through an irreducible complex phase in the weak-interaction couplings of quarks. A broad class of SM extensions allows for additional sources of \CP\ violation, which, if observed, could provide indirect indications of unknown particles or interactions. To date, \CP\ violation has been established in transitions of strange and bottom hadrons, with effects consistent with the SM predictions~\cite{hfag,Antonelli:2009ws}. Studies of \CP\ violation in the interactions of charm quarks offer a unique probe for non-SM physics. Charm transitions are complementary to the processes involving $K$ and $B$ mesons in that heavy up-type quarks (charge +2/3) are present in the initial state. Therefore, measurements of \CP\ violation in charm probe the presence of down-type (charge $-1/3$) non-SM physics through charged-current couplings~\cite{theory}. Because charm transitions are well described by the physics of the first two quark generations, \CP-violating effects are expected not to exceed $\mathcal{O}(10^{-2})$ in the SM~\cite{theory}. Indeed, no \CP\ violation has been experimentally established yet in charm-quark dynamics~\cite{hfag}.\par
%
Decay-time-dependent rate asymmetries of Cabibbo-suppressed decays into \CP\ eigenstates, such as $D \to h^+h^-$, where $D$ indicates a $D^0$ or $\overline{D}^0$ meson, and $h$ a $K$ or $\pi$ meson, are among the most sensitive probes for \CP\ violation in this sector~\cite{SCS-decays}. Such asymmetries,
\begin{equation}\label{eq:acp}
\Acp(t) = \frac{d\Gamma(D^0\to h^+h^-)/dt - d\Gamma(\Dbar^0\to h^+h^-)/dt}{d\Gamma(D^0\to h^+h^-)/dt+d\Gamma(\Dbar^0\to h^+h^-)/dt},
\end{equation}
\noindent probe non-SM physics contributions in the oscillation and  penguin transition amplitudes. Oscillations indicate $D^0$--$\Dbar^0$ transitions governed by the exchange of virtual heavy particles occurring before the decay. Penguin decays are second-order transitions mediated by an internal loop. Either amplitude may be affected by the exchange of non-SM particles, which could enhance the magnitude of the observed \CP\ violation with respect to the SM expectation. The asymmetry $\Acp(t)$ thus receives contributions from any difference between $D^0$ and $\Dbar^0$ decay amplitudes (direct \CP\ violation) and from any difference in oscillation probabilities between charm and anticharm mesons or interference between decays that follow, or not, an oscillation (indirect \CP\ violation). Because of the slow oscillation rate of charm mesons~\cite{hfag}, Eq.~\eqref{eq:acp} is approximated to first order as \cite{AcpCDF}
\begin{equation}\label{eq:acp3}
\Acp(t) \approx \Acp^{\rm{dir}}(h^+h^-) - \frac{t}{\tau}\ A_\Gamma(h^+h^-),
\end{equation}
\noindent where $t$ is the proper decay time 
and $\tau$ is the \CP-averaged $D$-meson lifetime~\cite{pdg}. The first term arises from direct \CP\ violation and depends 
on the decay mode; the second term is proportional to the asymmetry between the {\it effective} lifetimes $\hat{\tau}$ of anticharm and charm mesons,
\begin{equation*}
A_{\Gamma} = \frac{\hat{\tau}(\Dbar^0\to h^+h^-)-\hat{\tau}(D^0\to h^+h^-)}{\hat{\tau}(\Dbar^0\to h^+h^-)+\hat{\tau}(D^0\to h^+h^-)},
\end{equation*}
\noindent and is mostly due to indirect \CP\ violation~\cite{Gersabeck}. Effective lifetimes are defined as those resulting from a single-exponential fit of the time evolution of neutral meson decays that may undergo oscillations. In the SM, $A_{\Gamma}$ is universal for all final states with the same \CP-parity~\cite{Grossman:2006jg}, such as $K^+K^-$ and $\pi^+\pi^-$; contributions from non-SM processes may introduce channel-specific differences. Measurements have been reported from electron-positron collisions at the $\Upsilon(4S)$ resonance~\cite{BfactoriesAgamma} and from high-energy proton-proton collisions~\cite{LHCbAgamma}. All results are consistent with the hypothesis of \CP\ symmetry with $\mathcal{O}(10^{-3})$ uncertainties.\par
Any independent measurement of comparable precision further constrains the phenomenological bounds and may improve the knowledge of \CP\ violation in the charm sector. Decays $D \to h^+ h^-$ are well suited for a measurement of $A_{\Gamma}$ at the Collider Detector at Fermilab (CDF). Fully reconstructed final states provide a precise determination of the decay time, and large signal yields with moderate backgrounds allow for reduced systematic uncertainties.\par
%
In this paper, we report a measurement of \CP-violating asymmetries between the effective lifetimes of anticharm and charm mesons reconstructed in $D^0\to K^+ K^-$ and $D^0\to \pi^+\pi^-$ decays. We use the full data set from 1.96~\tev\ proton-antiproton collisions collected by the online event-selection system (trigger) on charged particles displaced from the primary collision and corresponding to $9.7$~fb$^{-1}$ of integrated luminosity. The analysis uses $D$-meson candidates produced in the decay of an identified $D^{*+}$ or $D^{*-}$ meson to determine whether the decaying state was initially produced as a $D^0$ or a $\Dbar^0$ meson. Flavor conservation in the strong-interaction processes $D^{*+} \to D^0 \pi_s^+$ and $D^{*-} \to \Dbar^0\pi_s^-$ allows identification of the initial flavor through the charge of the low-momentum $\pi$ meson (soft pion, $\pi_s$). Each decay-mode sample is divided into subsamples according to production flavor and decay time. In each subsample, a fit to the $D\pi_s^{\pm}$ mass distribution is used to determine the relative proportions of signal and background. These proportions are used to construct a background-subtracted distribution of the $D$ impact parameter, the minimum distance from the beam of the $D$ trajectory. This distribution is fit to identify  $D^{*\pm}$ mesons from $b$-hadron decays (secondary decays), whose observed decay-time distribution is biased by the additional decay length of the $b$-hadron, and to determine the yields of charm ($N_{D^{0}}$) and anticharm ($N_{\overline{D}^{0}}$) mesons directly produced in the $p\bar{p}$ collision (primary decays). The yields are combined into the asymmetry $A=(N_{D^0}-N_{\Dbar^0})/(N_{D^0}+N_{\Dbar^0})$, which is fit according to Eq.~\eqref{eq:acp3}. The slope yields $A_{\Gamma}$. The intercept determines the asymmetry at $t=0$, $A(0)$, which receives contributions from direct \CP\ violation and possible instrumental asymmetries. We check that the latter are constant in decay time using a low-background control sample of $13\times10^{6}$ $D^{*\pm}\to D(\to K^\mp\pi^\pm)\pi_s^\pm$ signal decays.  Sample selection, studies of background composition, and fit modeling follow previous measurements~\cite{AcpCDF,DeltaAcpCDF}.\par
%
The CDF II detector is a multipurpose magnetic spectrometer surrounded by calorimeters and muon detectors. The detector components relevant for this analysis are outlined as follows; a detailed description is in Ref.~\cite{AcpCDF}. A silicon microstrip vertex detector and a cylindrical drift chamber immersed in a $1.4$~T axial magnetic field allow reconstruction of charged-particle trajectories (tracks) in the pseudorapidity range $|\eta| < 1$. The vertex detector contains seven concentric layers of single- and double-sided silicon sensors at radii between 1.5 and 22~cm, each providing a position measurement with up to 15 (70) $\mu$m resolution in the azimuthal (proton-beam) direction~\cite{silicon}. The drift chamber has 96 measurement layers, between 40 and 137~cm in radius, organized into alternating axial and $\pm 2^{\circ}$ stereo superlayers~\cite{COT}. The component of a charged-particle momentum transverse to the beam ($p_T$) is determined with a resolution of $\sigma_{p_T}/p_T^2 \approx 0.07\%\ (\pgev)^{-1}$, corresponding to a typical mass resolution of $8~\massmev$ for a two-body charm-meson decay.
\par The data are collected by a three-level trigger. At level 1, custom hardware processors reconstruct tracks in the transverse plane of the drift chamber~\cite{XFT}. Two oppositely-charged particles are required, with reconstructed transverse momenta $p_{T} > 2~\pgev$, scalar sum $\sum p_{T} > 5.5~\pgev$, and azimuthal opening angle $\Delta \phi < 90^{\circ}$. At level 2, drift-chamber tracks are combined with silicon-detector hits and their impact parameters (transverse distances of closest approach to the beam line) are determined with 45~$\mu$m resolution (including the beam spread)~\cite{SVT} and required to be between 0.12 and 1.0 mm. A more stringent opening-angle requirement of $2^{\circ}<\Delta\phi<90^{\circ}$ is also applied. Each track pair is then used to form a $D$-meson candidate, whose flight distance in the transverse plane projected onto the transverse momentum ($L_{xy}$) is required to exceed $200~\mu$m. At level 3, the selection is reapplied on events fully reconstructed by an array of commercial processors.\par 
The offline reconstruction of signal candidates is solely based on tracking information, without using particle identification. Two tracks from oppositely-charged particles compatible with the trigger requirements are combined, with pion or kaon assignment, in a kinematic fit to a common decay vertex to form a $D$ candidate. A charged particle with $p_T>400~\pmev$ is associated with each $D$ candidate to form $D^{*\pm}$ candidates. We improve the reconstruction with respect to Ref.~\cite{DeltaAcpCDF} by using the position of the beam as a constraint in the fit of the $D^{*\pm}$ decay and retain only candidates with good fit quality. Since the beam position is determined more accurately than the trajectory of the soft pion, this provides a 25\% improvement in $D^{*\pm}$ mass resolution. 
Other offline requirements are based on a more accurate determination of the quantities used in the trigger and are detailed in Ref.~\cite{DeltaAcpCDF}. The $D \to K^+K^-$ and $D \to \pi^+\pi^-$ samples are separated by requiring the selected candidates to have the relevant $h^+h^-$ mass within 24 \massmev\ of the known $D$ mass, $m_{D}$~\cite{pdg}. 
\begin{figure*}[t]
\centering
\includegraphics[width=0.8\textwidth]{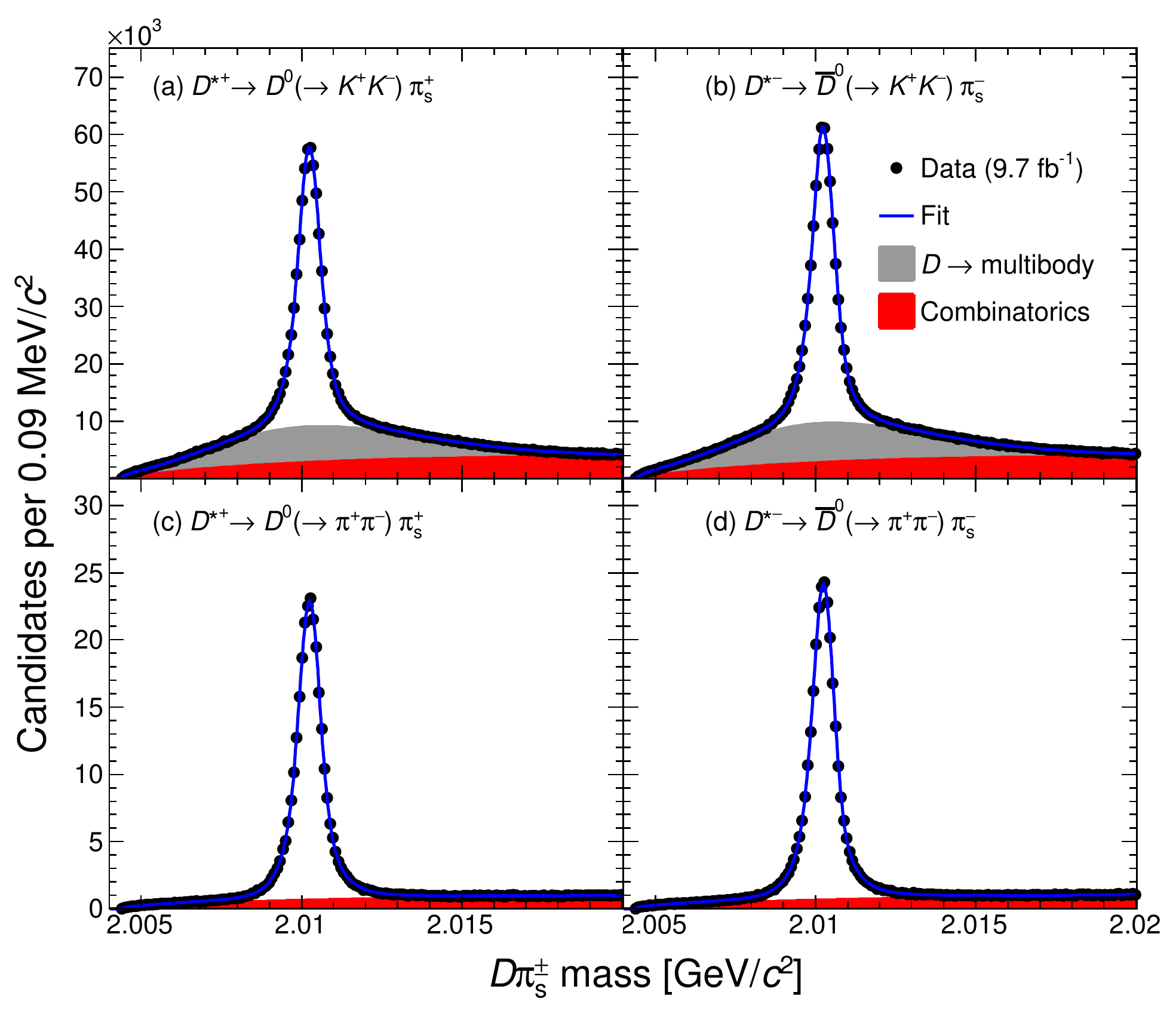}
\caption{Distributions of $D\pi_s^{\pm}$ mass with fit results overlaid for (a) the $D^0 \to K^+K^-$ sample,  (b) the $\overline{D}^0 \to K^+K^-$ sample, (c) the $D^0 \to \pi^+\pi^-$ sample, and (d) the $\overline{D}^0 \to \pi^+\pi^-$ sample.}\label{fig:mass}
\end{figure*}
We reconstruct $6.1\times 10^5$ $D^0 \to K^+K^-$,  $6.3\times 10^5$ $\overline{D}^0 \to K^+K^-$, $2.9\times 10^5$ $D^0 \to \pi^+\pi^-$, and $3.0\times 10^5$ $\overline{D}^0 \to \pi^+\pi^-$ signal decays (Fig.~\ref{fig:mass}). The composition of the $\pi^+\pi^-$ sample is dominated by the signal of $D^*$-tagged $D$ decays and a background of real $D$ decays associated with random pions or random combinations of three tracks (combinatorics). In the $K^+K^-$ sample, an additional background is contributed by misreconstructed multibody charm-meson decays, dominated by $D^0 \to h^-\pi^+ \pi^0$ and the  $D^0 \to h^- \ell^+ \nu_{\ell}$ contributions, where $\ell$ is a muon or an electron.\par
%
Each decay-mode sample is divided into charm and anticharm subsamples and into 30 bins of decay time between $0.15\tau$ and $20\tau$, chosen so that each contains approximately the same number of candidates. The $D$ decay time is determined as $t=L_{xy}m_{D}/p_T$, with approximately $0.2\tau$ resolution, independent of decay time. 
The observed decay-time distribution is biased by the trigger. The effect of the bias is assumed to be independent of the $D$-meson flavor and is accounted for when integrating Eq.~(\ref{eq:acp3}) over each decay-time bin.\par
Relative proportions between signal and background yields in the signal region are determined in each decay-time bin, and for each flavor, through $\chi^2$ fits of the $D\pi_{s}^{\pm}$ mass distributions. The $D\pi_s^{\pm}$ mass is calculated using the vector sum of the momenta of the three particles to determine the $D^{*\pm}$ momentum and the known $D$ and charged $\pi$-meson masses~\cite{pdg}. The signal shapes are determined from the sample of $D\to K^\mp\pi^\pm$ decays; the parameters of the background shapes~\cite{AcpCDF} are determined by the fit. All mass shapes are determined independently for each flavor and decay-time bin. The fit allows for asymmetries between combinatorial and misreconstructed background event yields, respectively, of the $D^{*+}$ and $D^{*-}$ samples. The resulting shapes and background proportions are used to derive signal-only distributions of the $D$-meson impact parameter in each bin and for each flavor.\par
\begin{figure}
\centering
\includegraphics[width=0.5\textwidth]{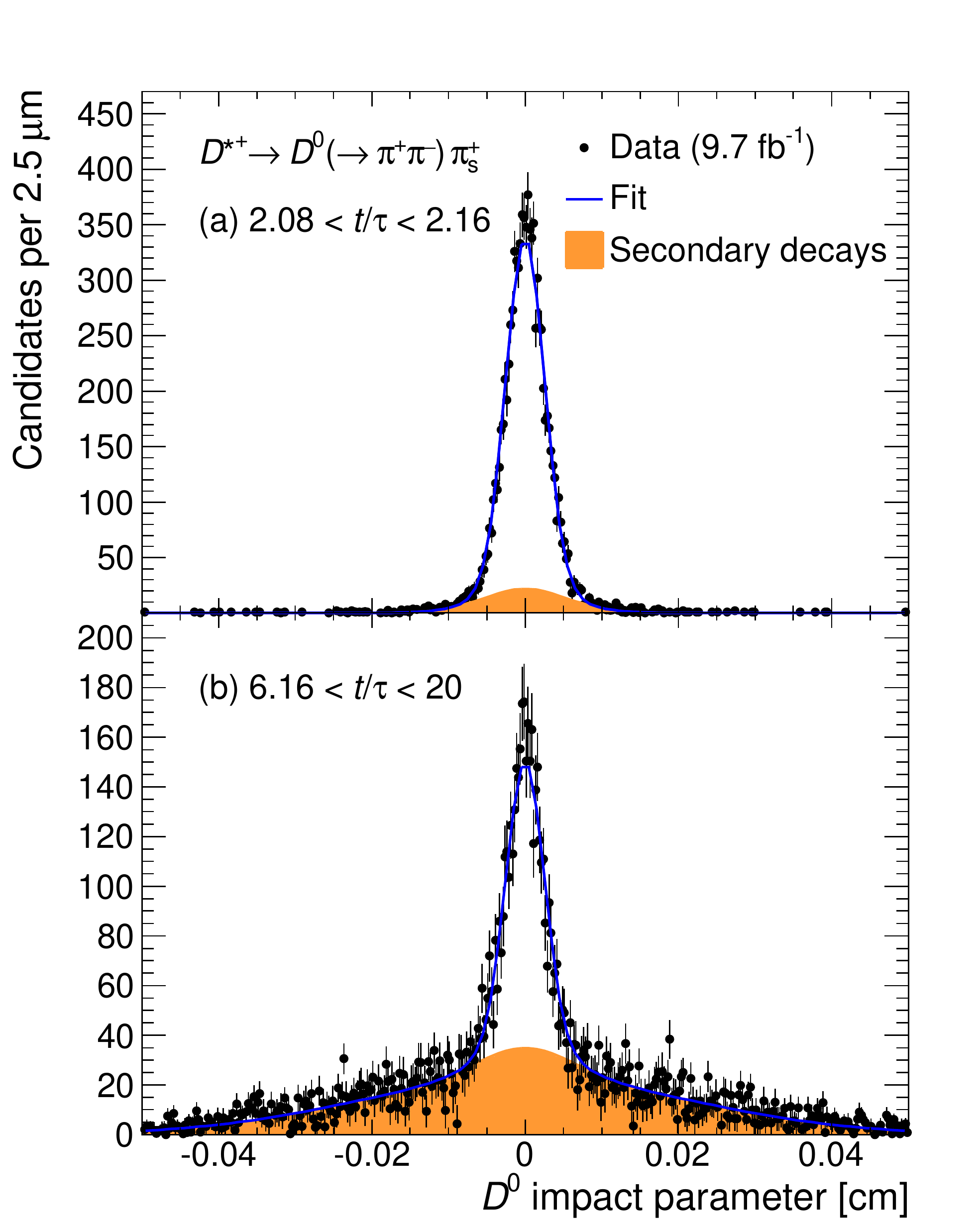}
\caption{Distributions of $D$-meson impact parameter with fit results overlaid for background-subtracted $D \to \pi^+\pi^-$ decays restricted to (a) the decay-time bin $2.08 < t/\tau < 2.16$ and (b) the decay-time bin $6.16 < t/\tau < 20$. Similar distributions are observed for $D \to K^+K^-$ decays.}\label{fig:ip}
\end{figure}

The impact parameter distributions of the sum of signal and background components are formed by restricting the analysis to candidates with $M(D\pi_s^\pm)$ within $2.4~\massmev$ of the known $D^{*\pm}$ mass~\cite{pdg}. From these, we subtract the impact parameter distribution of the background, derived from the $2.015 < M(D\pi^{\pm})<2.020~\massgev$ region for the $\pi^+\pi^-$ sample. The additional contamination from multibody decays in the $K^+K^-$ sample requires choosing a suitable sideband that contains the same admixture of combinatorial and misreconstructed backgrounds as that expected in the signal region. We select as background the candidates with $m_{D}-64~\massmev<M(K^+K^-)<m_{D}-40~\massmev$ and with $M(D\pi_s^\pm)$ within $2.4~\massmev$ of the known $D^{*\pm}$ mass. Checks on data show that the final results are robust against variations of these choices. We perform a $\chi^2$ fit of the background-subtracted impact-parameter distribution of $D$ candidates in each subsample of decay-time and flavor, using double-Gaussian models for both the primary and secondary components. Since we determine impact parameters using information associated with the $D$ decay only, the shapes of the impact-parameter distributions of $D^0$ and $\Dbar^0$ mesons are consistent. The parameters of the primary component are fixed in all fits. They are derived from fits of candidates in the first decay-time bin ($t/\tau<1.18$), where any bias from the $\mathcal{O}(\%)$ secondary contamination is negligible, as supported by repeating the fit using an alternative model derived from the second bin and observing no significant difference in the results. The parameters of the secondary component are determined by the fit independently for each decay-time bin. Example impact-parameter fits are shown in Fig.~\ref{fig:ip}. All mass and impact parameter fits show good agreement with data. Extreme variations of model parameters yield large changes in fit $\chi^2$ but negligible changes of the results.
\par
\begin{figure}
\centering
\includegraphics[width=0.5\textwidth]{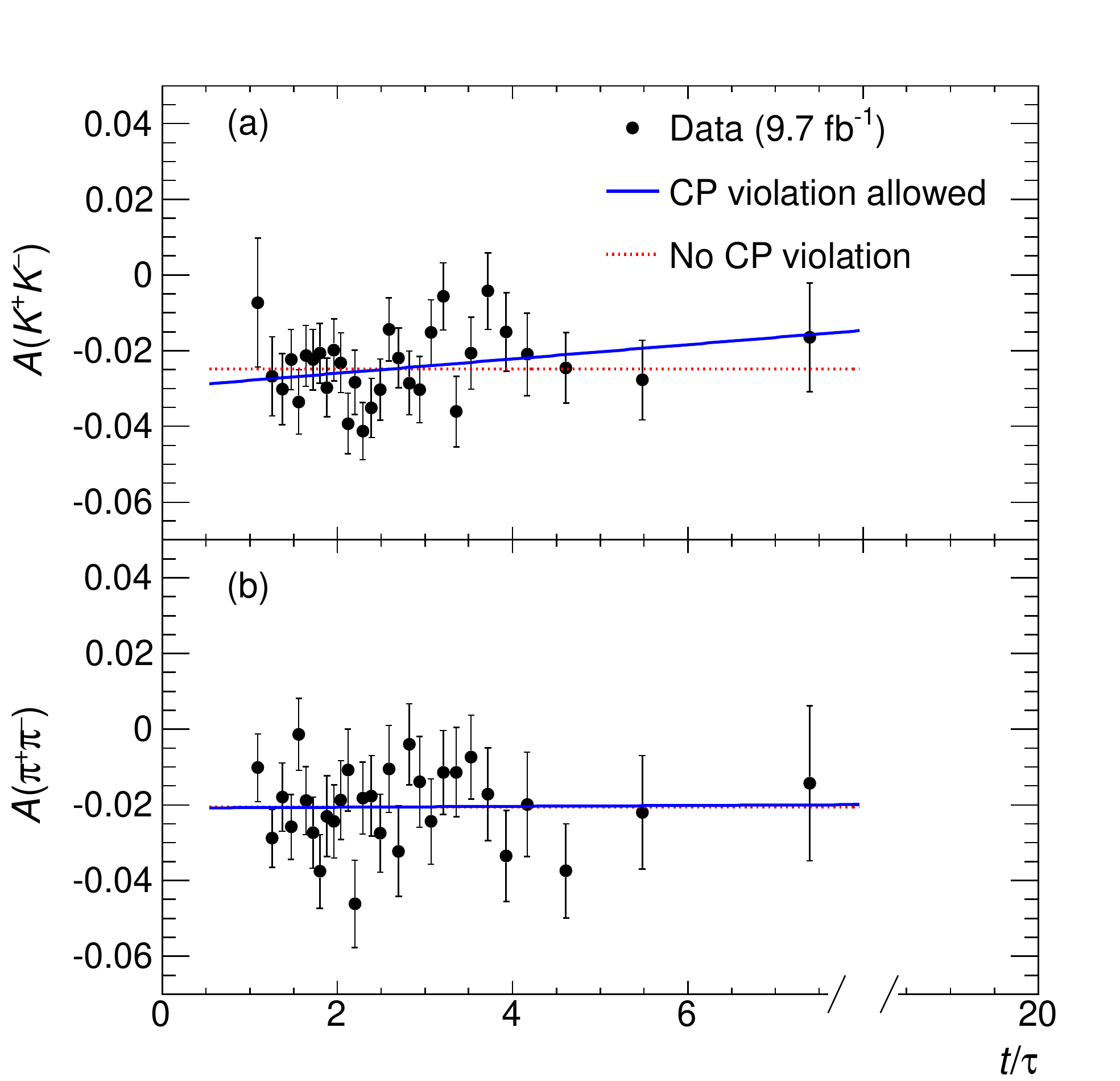}
\caption{Effective lifetime asymmetries as functions of decay time for the (a) $D \to K^+K^-$ and (b) $D \to\pi^+\pi^-$ samples. In each bin, the position of the data point corresponds to the average decay-time in that bin. Results of fits not allowing for (dotted line) and allowing for (solid line) \CP\ violation are overlaid.}\label{fig:Agamma}
\end{figure}

Final $\chi^2$ fits of the asymmetries between the resulting yields of primary charm and anticharm decays as functions of decay time are used to determine the values of $A_{\Gamma}$ in the two samples. The fits are shown in Fig.~\ref{fig:Agamma} and yield $A_{\Gamma}(K^+K^-) = \bigl(-0.19 \pm 0.15\stat\bigr)\%$ and $A_{\Gamma}(\pi^+\pi^-) = \bigl(-0.01 \pm 0.18\stat\bigr)\%$. The value of $\chi^2$ divided by the number of degrees of freedom is 28/28 in both fits. In both samples we observe $A(0) \approx -2\%$, due to the known detector-induced asymmetry in the soft-pion reconstruction efficiency~\cite{AcpCDF}. The independence of instrumental asymmetries from decay time is checked by performing the analysis on $D \to K^\mp\pi^\pm$ decays, where no indirect \CP\ violation occurs and instrumental asymmetries are larger due to the additional effect from the difference in interaction probability with matter of opposite-charge kaons; an asymmetry slope compatible with zero is found, $(-0.5 \pm 0.3)\times 10^{-3}$. The width of the impact-parameter distribution of primary $D$ mesons increases as a function of decay time, as predicted in simulation. This has no significant effect on $A_{\Gamma}$, as verified by repeating the measurement with a floating width that increases linearly with decay time.\par
The dominant systematic uncertainty in the measurement of $A_\Gamma(\pi^+\pi^-)$, arises from the contribution of $\pm 0.028\%$ from
the choice of the impact-parameter shape (single or double Gaussian function) of the secondary component whereas for $A_\Gamma(K^+K^-)$ this effect contributes a smaller uncertainty of $\pm 0.013\%$. The choice of the background sideband has a dominant effect in the $K^+K^-$ analysis ($\pm 0.038\%$) and a minor impact ($\pm 0.010\%$) on the $\pi^+\pi^-$ result. Other minor effects are associated with the uncertainty on the vertex-detector length-scale ($\pm 0.001\%$ to $\pm 0.002\%$); the neglected 0.93\% contamination of misreconstructed $K^-\pi^+$ decays in the $\pi^+\pi^-$ sample ($<0.001\%$); the neglected bin-by-bin migration due to the decay-time resolution ($<0.001\%$); and any possible fit biases ($<0.001\%$), probed by repeating the analysis on the $\pi^+\pi^-$ sample with random flavor assignment. \par
%
In summary, we measure the difference in effective lifetime between anticharm and charm mesons reconstructed in $D^0 \to K^+K^-$ and $D^0 \to \pi^+\pi^-$ decays using the full CDF data set. The final results,
\begin{align*}
A_{\Gamma}(K^+K^-) = (-0.19 \pm 0.15 \stat  \pm 0.04 \syst)\%,\\
A_{\Gamma}(\pi^+\pi^-) = (-0.01 \pm 0.18 \stat  \pm 0.03 \syst)\%,
\end{align*}
\noindent are consistent with the hypothesis of \CP\ symmetry. Their combination yields $A_\Gamma = \bigl(-0.12 \pm 0.12)\%$, assuming that uncertainties are uncorrelated. The results are consistent with the current best determinations \cite{BfactoriesAgamma,LHCbAgamma} and improve the global constraints on indirect \CP\ violation in charm-meson dynamics.  

\input{acknowledgements}

\end{document}

%% file: author.tex
\affiliation{Institute of Physics, Academia Sinica, Taipei, Taiwan 11529, Republic of China}
\affiliation{Argonne National Laboratory, Argonne, Illinois 60439, USA}
\affiliation{University of Athens, 157 71 Athens, Greece}
\affiliation{Institut de Fisica d'Altes Energies, ICREA, Universitat Autonoma de Barcelona, E-08193, Bellaterra (Barcelona), Spain}
\affiliation{Baylor University, Waco, Texas 76798, USA}
\affiliation{Istituto Nazionale di Fisica Nucleare Bologna, \ensuremath{^{ii}}University of Bologna, I-40127 Bologna, Italy}
\affiliation{University of California, Davis, Davis, California 95616, USA}
\affiliation{University of California, Los Angeles, Los Angeles, California 90024, USA}
\affiliation{Instituto de Fisica de Cantabria, CSIC-University of Cantabria, 39005 Santander, Spain}
\affiliation{Carnegie Mellon University, Pittsburgh, Pennsylvania 15213, USA}
\affiliation{Enrico Fermi Institute, University of Chicago, Chicago, Illinois 60637, USA}
\affiliation{Comenius University, 842 48 Bratislava, Slovakia; Institute of Experimental Physics, 040 01 Kosice, Slovakia}
\affiliation{Joint Institute for Nuclear Research, RU-141980 Dubna, Russia}
\affiliation{Duke University, Durham, North Carolina 27708, USA}
\affiliation{Fermi National Accelerator Laboratory, Batavia, Illinois 60510, USA}
\affiliation{University of Florida, Gainesville, Florida 32611, USA}
\affiliation{Laboratori Nazionali di Frascati, Istituto Nazionale di Fisica Nucleare, I-00044 Frascati, Italy}
\affiliation{University of Geneva, CH-1211 Geneva 4, Switzerland}
\affiliation{Glasgow University, Glasgow G12 8QQ, United Kingdom}
\affiliation{Harvard University, Cambridge, Massachusetts 02138, USA}
\affiliation{Division of High Energy Physics, Department of Physics, University of Helsinki, FIN-00014, Helsinki, Finland; Helsinki Institute of Physics, FIN-00014, Helsinki, Finland}
\affiliation{University of Illinois, Urbana, Illinois 61801, USA}
\affiliation{The Johns Hopkins University, Baltimore, Maryland 21218, USA}
\affiliation{Institut f\"{u}r Experimentelle Kernphysik, Karlsruhe Institute of Technology, D-76131 Karlsruhe, Germany}
\affiliation{Center for High Energy Physics: Kyungpook National University, Daegu 702-701, Korea; Seoul National University, Seoul 151-742, Korea; Sungkyunkwan University, Suwon 440-746, Korea; Korea Institute of Science and Technology Information, Daejeon 305-806, Korea; Chonnam National University, Gwangju 500-757, Korea; Chonbuk National University, Jeonju 561-756, Korea; Ewha Womans University, Seoul, 120-750, Korea}
\affiliation{Ernest Orlando Lawrence Berkeley National Laboratory, Berkeley, California 94720, USA}
\affiliation{University of Liverpool, Liverpool L69 7ZE, United Kingdom}
\affiliation{University College London, London WC1E 6BT, United Kingdom}
\affiliation{Centro de Investigaciones Energeticas Medioambientales y Tecnologicas, E-28040 Madrid, Spain}
\affiliation{Massachusetts Institute of Technology, Cambridge, Massachusetts 02139, USA}
\affiliation{University of Michigan, Ann Arbor, Michigan 48109, USA}
\affiliation{Michigan State University, East Lansing, Michigan 48824, USA}
\affiliation{Institution for Theoretical and Experimental Physics, ITEP, Moscow 117259, Russia}
\affiliation{University of New Mexico, Albuquerque, New Mexico 87131, USA}
\affiliation{The Ohio State University, Columbus, Ohio 43210, USA}
\affiliation{Okayama University, Okayama 700-8530, Japan}
\affiliation{Osaka City University, Osaka 558-8585, Japan}
\affiliation{University of Oxford, Oxford OX1 3RH, United Kingdom}
\affiliation{Istituto Nazionale di Fisica Nucleare, Sezione di Padova, \ensuremath{^{jj}}University of Padova, I-35131 Padova, Italy}
\affiliation{University of Pennsylvania, Philadelphia, Pennsylvania 19104, USA}
\affiliation{Istituto Nazionale di Fisica Nucleare Pisa, \ensuremath{^{kk}}University of Pisa, \ensuremath{^{ll}}University of Siena, \ensuremath{^{mm}}Scuola Normale Superiore, I-56127 Pisa, Italy, \ensuremath{^{nn}}INFN Pavia, I-27100 Pavia, Italy, \ensuremath{^{oo}}University of Pavia, I-27100 Pavia, Italy}
\affiliation{University of Pittsburgh, Pittsburgh, Pennsylvania 15260, USA}
\affiliation{Purdue University, West Lafayette, Indiana 47907, USA}
\affiliation{University of Rochester, Rochester, New York 14627, USA}
\affiliation{The Rockefeller University, New York, New York 10065, USA}
\affiliation{Istituto Nazionale di Fisica Nucleare, Sezione di Roma 1, \ensuremath{^{pp}}Sapienza Universit\`{a} di Roma, I-00185 Roma, Italy}
\affiliation{Mitchell Institute for Fundamental Physics and Astronomy, Texas A\&M University, College Station, Texas 77843, USA}
\affiliation{Istituto Nazionale di Fisica Nucleare Trieste, \ensuremath{^{qq}}Gruppo Collegato di Udine, \ensuremath{^{rr}}University of Udine, I-33100 Udine, Italy, \ensuremath{^{ss}}University of Trieste, I-34127 Trieste, Italy}
\affiliation{University of Tsukuba, Tsukuba, Ibaraki 305, Japan}
\affiliation{Tufts University, Medford, Massachusetts 02155, USA}
\affiliation{University of Virginia, Charlottesville, Virginia 22906, USA}
\affiliation{Waseda University, Tokyo 169, Japan}
\affiliation{Wayne State University, Detroit, Michigan 48201, USA}
\affiliation{University of Wisconsin, Madison, Wisconsin 53706, USA}
\affiliation{Yale University, New Haven, Connecticut 06520, USA}

\author{T.~Aaltonen}
\affiliation{Division of High Energy Physics, Department of Physics, University of Helsinki, FIN-00014, Helsinki, Finland; Helsinki Institute of Physics, FIN-00014, Helsinki, Finland}
\author{S.~Amerio\ensuremath{^{jj}}}
\affiliation{Istituto Nazionale di Fisica Nucleare, Sezione di Padova, \ensuremath{^{jj}}University of Padova, I-35131 Padova, Italy}
\author{D.~Amidei}
\affiliation{University of Michigan, Ann Arbor, Michigan 48109, USA}
\author{A.~Anastassov\ensuremath{^{v}}}
\affiliation{Fermi National Accelerator Laboratory, Batavia, Illinois 60510, USA}
\author{A.~Annovi}
\affiliation{Laboratori Nazionali di Frascati, Istituto Nazionale di Fisica Nucleare, I-00044 Frascati, Italy}
\author{J.~Antos}
\affiliation{Comenius University, 842 48 Bratislava, Slovakia; Institute of Experimental Physics, 040 01 Kosice, Slovakia}
\author{G.~Apollinari}
\affiliation{Fermi National Accelerator Laboratory, Batavia, Illinois 60510, USA}
\author{J.A.~Appel}
\affiliation{Fermi National Accelerator Laboratory, Batavia, Illinois 60510, USA}
\author{T.~Arisawa}
\affiliation{Waseda University, Tokyo 169, Japan}
\author{A.~Artikov}
\affiliation{Joint Institute for Nuclear Research, RU-141980 Dubna, Russia}
% ...

\collaboration{CDF Collaboration}
\altaffiliation[With visitors from]{
\ensuremath{^{a}}University of British Columbia, Vancouver, BC V6T 1Z1, Canada,
\ensuremath{^{b}}Istituto Nazionale di Fisica Nucleare, Sezione di Cagliari, 09042 Monserrato (Cagliari), Italy,
\ensuremath{^{c}}University of California Irvine, Irvine, CA 92697, USA,
\ensuremath{^{d}}Institute of Physics, Academy of Sciences of the Czech Republic, 182~21, Czech Republic,
\ensuremath{^{e}}CERN, CH-1211 Geneva, Switzerland,
\ensuremath{^{f}}Cornell University, Ithaca, NY 14853, USA,
\ensuremath{^{g}}University of Cyprus, Nicosia CY-1678, Cyprus,
\ensuremath{^{h}}Office of Science, U.S. Department of Energy, Washington, DC 20585, USA,
\ensuremath{^{i}}University College Dublin, Dublin 4, Ireland,
\ensuremath{^{j}}ETH, 8092 Z\"{u}rich, Switzerland,
\ensuremath{^{k}}University of Fukui, Fukui City, Fukui Prefecture, Japan 910-0017,
\ensuremath{^{l}}Universidad Iberoamericana, Lomas de Santa Fe, M\'{e}xico, C.P. 01219, Distrito Federal,
\ensuremath{^{m}}University of Iowa, Iowa City, IA 52242, USA,
\ensuremath{^{n}}Kinki University, Higashi-Osaka City, Japan 577-8502,
\ensuremath{^{o}}Kansas State University, Manhattan, KS 66506, USA,
\ensuremath{^{p}}Brookhaven National Laboratory, Upton, NY 11973, USA,
\ensuremath{^{q}}Queen Mary, University of London, London, E1 4NS, United Kingdom,
\ensuremath{^{r}}University of Melbourne, Victoria 3010, Australia,
\ensuremath{^{s}}Muons, Inc., Batavia, IL 60510, USA,
\ensuremath{^{t}}Nagasaki Institute of Applied Science, Nagasaki 851-0193, Japan,
\ensuremath{^{u}}National Research Nuclear University, Moscow 115409, Russia,
\ensuremath{^{v}}Northwestern University, Evanston, IL 60208, USA,
\ensuremath{^{w}}University of Notre Dame, Notre Dame, IN 46556, USA,
\ensuremath{^{x}}Universidad de Oviedo, E-33007 Oviedo, Spain,
\ensuremath{^{y}}CNRS-IN2P3, Paris, F-75205 France,
\ensuremath{^{z}}Universidad Tecnica Federico Santa Maria, 110v Valparaiso, Chile,
\ensuremath{^{aa}}The University of Jordan, Amman 11942, Jordan,
\ensuremath{^{bb}}Universite catholique de Louvain, 1348 Louvain-La-Neuve, Belgium,
\ensuremath{^{cc}}University of Z\"{u}rich, 8006 Z\"{u}rich, Switzerland,
\ensuremath{^{dd}}Massachusetts General Hospital, Boston, MA 02114 USA,
\ensuremath{^{ee}}Harvard Medical School, Boston, MA 02114 USA,
\ensuremath{^{ff}}Hampton University, Hampton, VA 23668, USA,
\ensuremath{^{gg}}Los Alamos National Laboratory, Los Alamos, NM 87544, USA,
\ensuremath{^{hh}}Universit\`{a} degli Studi di Napoli Federico I, I-80138 Napoli, Italy
}
\noaffiliation

%% file: acknowledgements.tex
We thank the Fermilab staff and the technical staffs of the
participating institutions for their vital contributions. This work
was supported by the U.S. Department of Energy and National Science
Foundation; the Italian Istituto Nazionale di Fisica Nucleare; the
Ministry of Education, Culture, Sports, Science and Technology of
Japan; the Natural Sciences and Engineering Research Council of
Canada; the National Science Council of the Republic of China; the
Swiss National Science Foundation; the A.P. Sloan Foundation; the
Bundesministerium f\"ur Bildung und Forschung, Germany; the Korean
World Class University Program, the National Research Foundation of
Korea; the Science and Technology Facilities Council and the Royal
Society, United Kingdom; the Russian Foundation for Basic Research;
the Ministerio de Ciencia e Innovaci\'{o}n, and Programa
Consolider-Ingenio 2010, Spain; the Slovak R\&D Agency; the Academy
of Finland; the Australian Research Council (ARC); and the EU community
Marie Curie Fellowship Contract No. 302103.